\documentstyle[preprint,aps,prl,epsf]{revtex}   
\include{amsmath}

\begin{document}
\title{{\bf Nanoscale fluid flows in the vicinity of patterned surfaces}}

\author
{Marek Cieplak,$^{1}$ Joel Koplik,$^{2}$ Jayanth R. Banavar$^{3}$\\}

\address{
\normalsize{$^{1}$Institute of Physics, Polish Academy of Sciences,}\\
\normalsize{Aleja Lotnik\'ow 32/46, 02-668 Warsaw, Poland}\\
\normalsize{$^{2}$Benjamin Levich Institute and Department of Physics,}\\
\normalsize{City College of the City University of New York, NY 10031, USA}\\
\normalsize{$^{3}$Department of Physics, 104 Davey Laboratory, The Pennsylvania
State University,}\\
\normalsize{University Park, Pennsylvania 16802, USA}\\
}


\maketitle

\date{}
\vskip 40pt


\vspace*{1cm}


\baselineskip24pt

\begin{abstract}
Molecular dynamics simulations of dense and rarefied fluids
comprising small chain molecules in chemically patterned
nano-channels predict a novel switching from Poiseuille to
plug flow along the channel. We also demonstrate behavior
akin to the lotus effect for a nanodrop on a chemically patterned
substrate.
Our results show that one can control and exploit the behavior of fluids
at the nanoscale using chemical patterning.
\end{abstract}

\newpage

Recent advances in fabrication  techniques 
\cite{PhysicsTodayVieu,GillmorDarRus}   
have spawned the field of nanofluidics where minute amounts of fluid are 
contained and controlled for practical use.
Here we study fluids  
of chain molecules flowing in chemically patterned nano-channels 
\cite{Laibinis,Huskens}     
or placed on a chemically patterned substrate using molecular dynamics
simulations. We demonstrate existence of novel flow patterns
and find a nanoscale version of the lotus 
effect \cite{Barthlott}.\\  

Understanding the behavior of fluids sometimes requires information
at the molecular scale, e.g. about the 
boundary conditions \cite{Maxwell}.
The continuum description itself may break down as in a rarified
fluid in the Knudsen regime \cite{Kennard,knudsenprl}  
in which the mean free path is large.
Computer simulations bridge
molecular scale physics and the continuum 
behavior \cite{Alder}    
and are ideally suited to study nanosized systems.
Understanding of liquid and gas flows in nanoscale devices is of great
technological interest \cite{PhysicsTodayVieu,Squires} 
and requires a molecular scale analysis 
both because
the boundary conditions determined by the
fluid-solid interactions are all-important and 
because of dilution.\\

What are the effects of wettability in 
the transport of fluids
in narrow channels? We consider a fluid comprising small
chain molecules in a
nanoscale channel with a sharp step in the 
solid-fluid intermolecular interactions.  
The nano-channel is bounded by two parallel walls in the $x-y$ plane.
Periodic boundary conditions are imposed
along the $x$ and $y$ directions. Fluid atoms at 
a distance $r$
interact with the Lennard-Jones
potential $V_{LJ}(r)= 4 \epsilon \left[ (\frac{r}{\sigma})^{-12} \;-\;
(\frac{r}{\sigma})^{-6} \right] \;$, where
$\sigma$ is the size of the repulsive core. 
The potential is truncated at $2.2 \sigma$ and shifted \cite{Robbins}.
The atoms are partitioned into chains of length $n$=10. The
tethering within the chains is provided
by the FENE potential \cite{Kroger}, 
$V_{FENE}=-(\kappa r_0^2 /2) \,
log[1-(r/r_0)^2]$, where $\kappa =30\epsilon$ and
$r_0=1.5\sigma$ with neighboring atoms of a chain separated by a distance
of around $\sigma$. The system is
maintained at a temperature $T=1.6\epsilon/k_B$ --  
above the liquid-gas coexistence region of this model.
The Langevin thermostat \cite{Allen} 
is used with the noise applied in all
directions during equilibration and 
in the $y$-direction afterwords. 
The radius of gyration of a
single molecule in the absence of flow is typically less than $1.7
\sigma$. The 
time unit, $\tau=
\sqrt{m\sigma ^2/\epsilon}$, where $m$ is the mass of the atom is of order
of 1 ps for typical fluids. The flow is
implemented through   
a "gravitational" acceleration of 0.01 $\sigma
/\tau ^2$ in the positive $x$-direction.
The walls are constructed from two [001] planes of an fcc lattice with a
lattice constant of $0.85 \sigma $. The
wall atoms are tethered to 
lattice sites by a harmonic spring with a
large spring constant. 
The fluid atoms were confined to a volume of
$26.2 \sigma \times 5.1
\sigma$ in the $x-y$ plane and $L_0 = 12.75 \sigma$ between the inner
faces of the walls. This corresponds to a
channel width of around $4.3$ nm for $\sigma$=3.4 $\AA$, corresponding to liquid
argon.\\

The wall-fluid interactions follow a Lennard-Jones potential
$V_{wf}(r)= 16 \epsilon \left[
(r/\sigma)^{-12} \;-\; A \; (r/\sigma)^{-6} \right] \;$
as in refs. \cite{knudsenprl} and \cite{Barrat}.
The parameter $A$ determines the wetting
properties of the wall and is varied between 1 and 0 corresponding to
attractive and repulsive walls respectively.
We assigned $A$=1 to the left half of the channel walls
and $A$=0 to the right half. Such a
step-wise variation can be accomplished experimentally by coating surfaces
with two different kinds of molecules \cite{Huskens}.
The number of fluid atoms 
is equal to $1200$ and
$240$ for the dense and rarified fluids
respectively. In the latter case, only two chains, on average, 
participate in a
ballistic motion at a given instant and the 
other chains coat the
attractive walls. The equations of motion are integrated
using a fifth-order predictor-corrector algorithm \cite{Allen}.
The spatial averaging is done  
in slabs of width $\sigma /4$ along the $z$- and $x$-axes. 
The rarified
fluid data are time averaged over 1 million $\tau _0$ to improve the
signal to noise ratio. The 
error bars are of the size of the data points.\\

The sharp step in the solid-fluid intermolecular interactions
leads to a jump in the
wettability, resulting in fluid adsorption
and layering on the attractive walls and a cushion of empty space
near the repulsive walls (Figure 1(a)). The
density profiles of the fluid 
near the walls (Figure 1(d))
play a key role in determining the flow
profile. 
In contrast to purely repulsive walls \cite{knudsenprl}, the inclusion of
regions of the channel wall with attractive interactions results in a
flow of the chain-molecule fluid achieving a steady state
independent of fluid density, rather than being accelerated without bounds.
The attractive walls provide a
mechanism for energy dissipation which is 
absent when walls are all repulsive.
(Interestingly, monatomic fluids do
reach a steady state 
in the presence of such walls).
For the flow rates considered here ($Re \sim 3$ and 8 in the
dense and rarefied cases respectively), the conformations of
molecules are fairly isotropic 
at the center of the channel but elongated along the flow direction 
at the attractive walls. More
strikingly, the character of the flow itself changes 
depending on the location of the channel and the nature of the
wall-fluid interactions.  Furthermore, the velocity fields are 
distinct for the dense fluid and in the
Knudsen regime (Figures 1(b) and 1(c)).\\ 

The velocity profiles (Figure 2(a)) of the nanoflows exhibit the expected
Poiseuelle character in the region of the
channel with attractive walls but one obtains a plug flow in the region
bounded by purely repulsive walls. Note that
the speed of the plug flow is smaller than the peak velocity of the
Poiseuelle flow in the dense fluid case while
the opposite behavior holds for the rarified fluid, underscoring the
important role played by fluid-fluid
interactions in controlling the dissipation. Indeed, in the Knudsen
regime, the fluid molecules in the region
bounded by repulsive walls undergo free acceleration with a 
moderating dissipation occurring only when they
pass through the region with attractive walls, resulting in a stationary
state. 
Figure 2(b) shows the slip length \cite{Kennard,knudsenprl,Schmatko}
and 2(c) the maximum velocity in different parts of the chemically patterned channel.\\

We have also studied channels that are patterned geometrically
with homogeneous attractive wall-fluid interactions. The extra space in
the wider fragments of the channel gets filled in by fluid molecules and the
resulting flow is similar to that in a
uniform width channel, 
underscoring the important
role played by the wetting heterogeneities compared to structural
patterning.  Earlier studies \cite{Charlaix} 
had considered a similar geometrically patterned channel but with 
uniformly repulsive walls leading to large slip and 
low flow resistance, akin to the
uniform channel case.\\

We turn now to the study of the 
behavior of a fluid nanodroplet on a substrate 
\cite{Querre}.    
The droplet (Figure 3) is made of 1800 fluid atoms resting on  
the bottom wall of the channel. The reduced temperature is chosen to be
$0.8$ so that 
the vapor pressure is essentially zero. 
The shape of the droplet on a surface is determined by the balance 
between the interfacial energies of the interfaces involved.
In our simulations there is a minimum wall-fluid
attraction, $A$, that is required to hold the droplet on the substrate.
This 
value, found to  
be around $A= 3/8$, separates two cases -- for $A> 3/8$, the droplet stays
on the substrate and otherwise it detaches.  
Our first set of runs were
carried out with a homogeneous substrate
characterized by this threshold value of $A$. The contact angle that one
obtains is 
around $110^{\rm o} \pm 10^{\rm o}$. \\

In order to assess the effects of the chemical patterning, we considered a modified
substrate in which the previous value of $A$ was restricted to
periodically repeated square patches placed in the background  
with an even smaller value of $A = 1/4$. The ratio of the patch size
(3.4 $\sigma$) to the droplet diameter (16 $\sigma$) and the 
strenght of the interaction with the background
were chosen 
so that the static drop was 
barely attached to the
substrate. 
The 
pattern leads to an increase in the contact angle (to $130 ^{\rm o}$ $\pm 10^{\rm o}$) 
without compromising the stability of the fluid on the substrate.\\

In the presence of a 
lateral push due to a small applied
acceleration 
the drop on the homogeneous
substrate has insignificant  
center-of-mass motion 
whereas
the drop on the patterned substrate readily
responds to the push with a combination of
rolling and diffusive motions (Figure 4). This behavior is 
akin to the 
lotus effect \cite{Barthlott}  
but on a nanoscale. The lotus leaf is self-cleansing because water droplets 
roll off easily and collect dirt, due to an interplay
between its rough microstructure and its
tendency to repel the water. In our case, a similar effect at the
nanoscale occurs even for 
a smooth substrate but with heterogenous wetting properties.\\

We have shown that molecular dynamics simulations can be used
to elucidate boundary conditions,
flow profiles in a patterned channel, and an analog of the lotus effect,
demonstrating 
novel behaviors that one may observe in the nanoworld.  
Controlling and harnessing fluids at the nanoscale
through chemical patterning may offer new opportunities in science and
technology.\\

We would like to acknowledge stimulating discussions with Amos Maritan and 
Giampaolo Mistura.
This work was supported by  Ministry of Science in Poland
(Grant No. 2P03B-03225), the  NASA Exploration Systems Mission Directorate, 
and by the European program IP NaPa through Warsaw
University of Technology.

\newpage
\begin{center}
{\bf Figure captions}
\end{center}

\vskip 0.5cm

\noindent
{\bf Fig. 1.} Molecular configuration, velocity fields, and density
profiles of fluids  
flowing through a patterned nano-channel. 
{\bf 1(a)}: A steady state snapshot of the dense
system of 120 chain molecules 
in an $x-z$ projection.
{\bf 1(b) and 1(c)}: The
velocity fields in the $x-z$ plane for the dense and
rarefied systems respectively. The 
darker vectors represent velocities larger than the average speed at the
center point of the system.  
In 1(b), the channel effectively widens leading to lower speeds on the
right. 
In 1(c), the speed is higher in the repulsive
region in spite of the increased effective width. 
{\bf 1(d)}: The density profiles for the dense (top) and rarefied (bottom) fluids.
Three locations along the
channel are shown for each case: the
dotted, broken, and solid lines are for $x/\sigma$ around 5.6 (denoted by
A), 13.8 (B) and 21 (C) corresponding to
attractive walls, the transition region, 
and repulsive walls respectively.
The atoms of the inner walls are at the edges of the figure. In the wetting
region, two monolayers   
form and make the effective width narrower.
In the non-wetting region, the adsorption is absent thus
the effective mid-point density
falls on moving from A to C.
This decrease (not
discernible in Figure 1(d)) 
is inconsequential for the dense case but explains the
anomalous behavior of the velocity field in the rarified case.\\

\noindent
{\bf Fig. 2.} Nano-flow atributes for the dense (left panels)
and rarefied (right panels) regimes.
{\bf 2(a)}: The velocity profiles for
the $x$ components of the flow velocity as
averaged in strips located at points A (asterisks), B (squares),
and C (circles) as in Figure 1(d).
The lines are 
guides to the eye.
One sees  
the Poiseuelle parabolic flow at A and B 
changing to plug flow at C. 
Dissipation 
at the attractive 
walls leads to stationarity of the flow that is missing for purely
repulsive walls.
{\bf 2(b)}: The slip length, $\zeta$, along the length of the channel. The arrow
shows the location of the switch-over in wettability.
$\zeta$ is obtained as a linear
extrapolation of the flow 
profile and 
is a measure of the distance from the wall at which the
extrapolated velocity vanishes.  
Positive (negative) 
$\zeta$ indicates that the 
extrapolated velocity vanishes
outside (inside) of the channel. 
$\zeta$ is negative for the attractive region
but becomes positive and large
when the fluid is within  
repulsive walls. 
The effective viscosity, as measured from the curvature of the velocity
profiles at the mid-point of the channel, is also a function 
of location along the channel and it follows the behavior of $\zeta$.
{\bf 2(c)}: The maximum
velocity, $v_{x,max}$, occuring half-way
between the walls, as a function of $x$.
It grows 
on crossing
the wettability step in the rarefied case
but decreases in the dense case.\\ 

\noindent
{\bf Fig. 3.} Snapshots of droplets on a solid substrate. The pictures
show a view (projection) from the top 
and a three-dimensional depiction of the droplet
in a side view (bottom).  
The left pictures refer to a patterned substrate with periodic
squares 
with $A=A_1=3/8$  superposed on a 
background of $A=A_2=1/4$ while  
those on the right depict a homogeneous surface with $A=3/8$ corresponding 
to partial wetting.
Smaller values of $A$ 
typically led to the detachment of
the droplet. 
The choice
of the $A_1$ and $A_2$ coefficients 
was dictated by the requirement of marginal attachment.
Note that the contact angle is larger for the patterned substrate
resulting in ''super-hydrophobicity".  
In some of our runs with the
patterned substrate, the droplet detaches 
when pushed laterally.\\

\noindent
{\bf Fig. 4.} The ``lotus" effect. Top: 
the time dependence of the longitudinal displacement of
the center of mass of the droplets 
pushed with
$g=0.002\tau ^2 /\sigma$ along the 
$x$-direction 
averaged over four runs.  
For the homogeneous substrate 
the droplet hardly moves but the motion is rapid
for the patterned substrate.
Bottom: the time dependence of  
height, $z$, above the substrate of droplet atoms
(two on the left, one on the right)
which are initially close to the surface of the droplet and the wall.
The nearly horizontal line indicates the location of the
center of mass 
of the droplet and provides a reference.
The pictures 
suggest rolling motion  \cite{Mahadevan} 
combined with diffusion 
for the patterned substrate (left) and  
pure diffusion for the homogenous substrate.\\

\newpage

\begin{figure*}
\epsfxsize=4.3in
\centerline{\epsffile{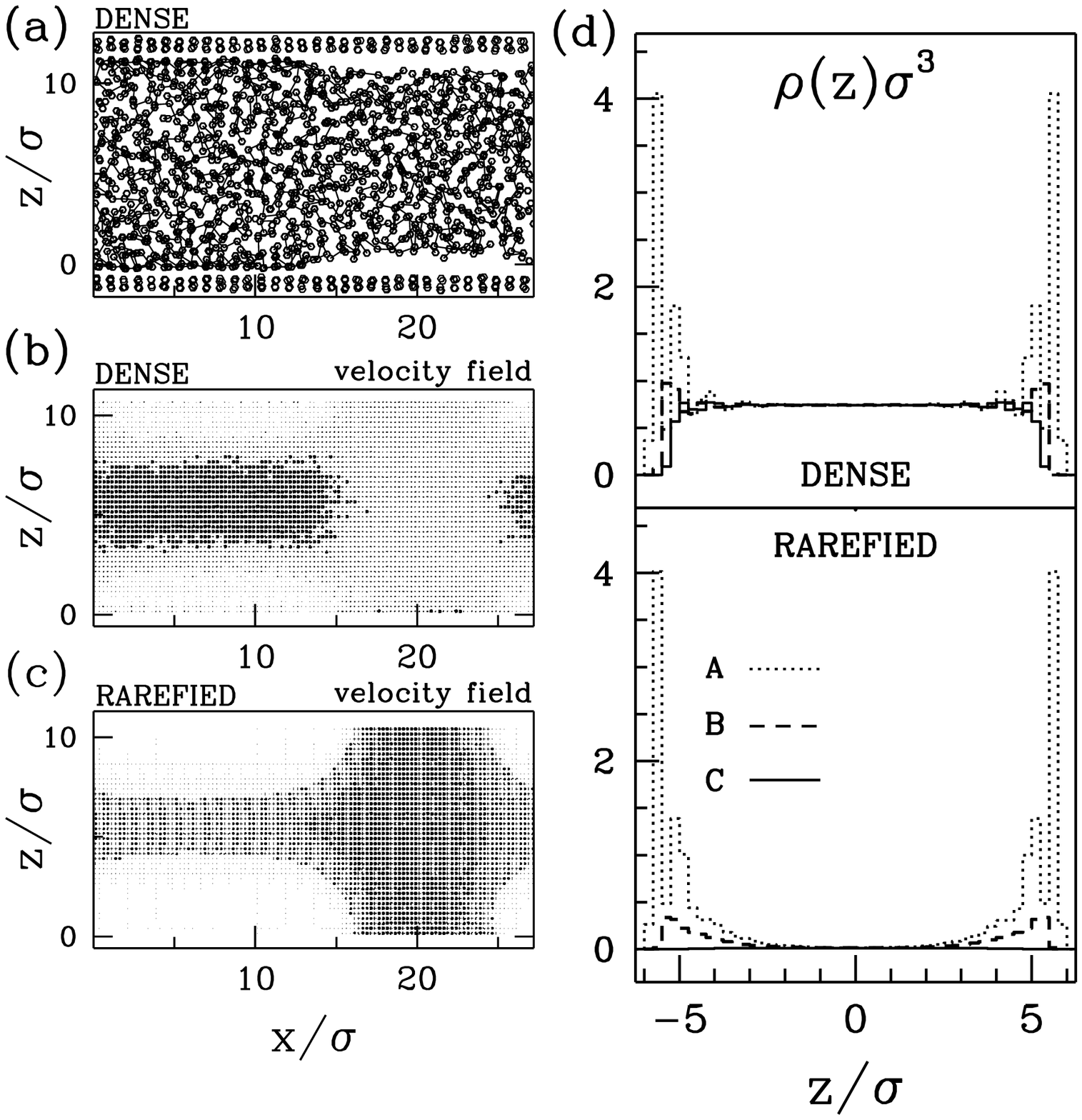}}
\vspace*{-3.5cm}
\caption{ }
\end{figure*}

\begin{figure*}
\epsfxsize=3.5in
\centerline{\epsffile{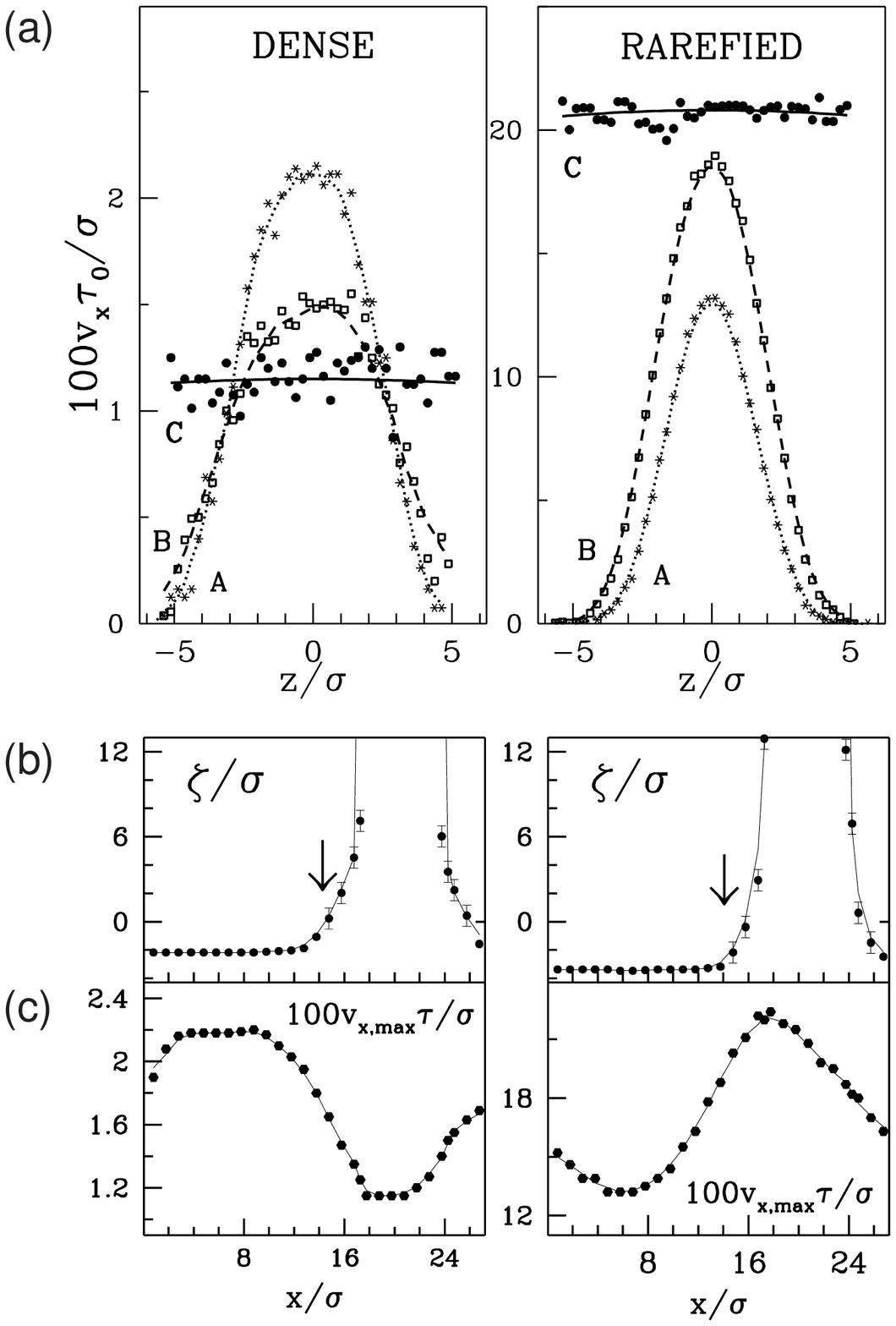}}
\vspace*{-0.4cm}
\caption{ }
\end{figure*}

\clearpage

\begin{figure*}
\epsfxsize=4.2in
\centerline{\epsffile{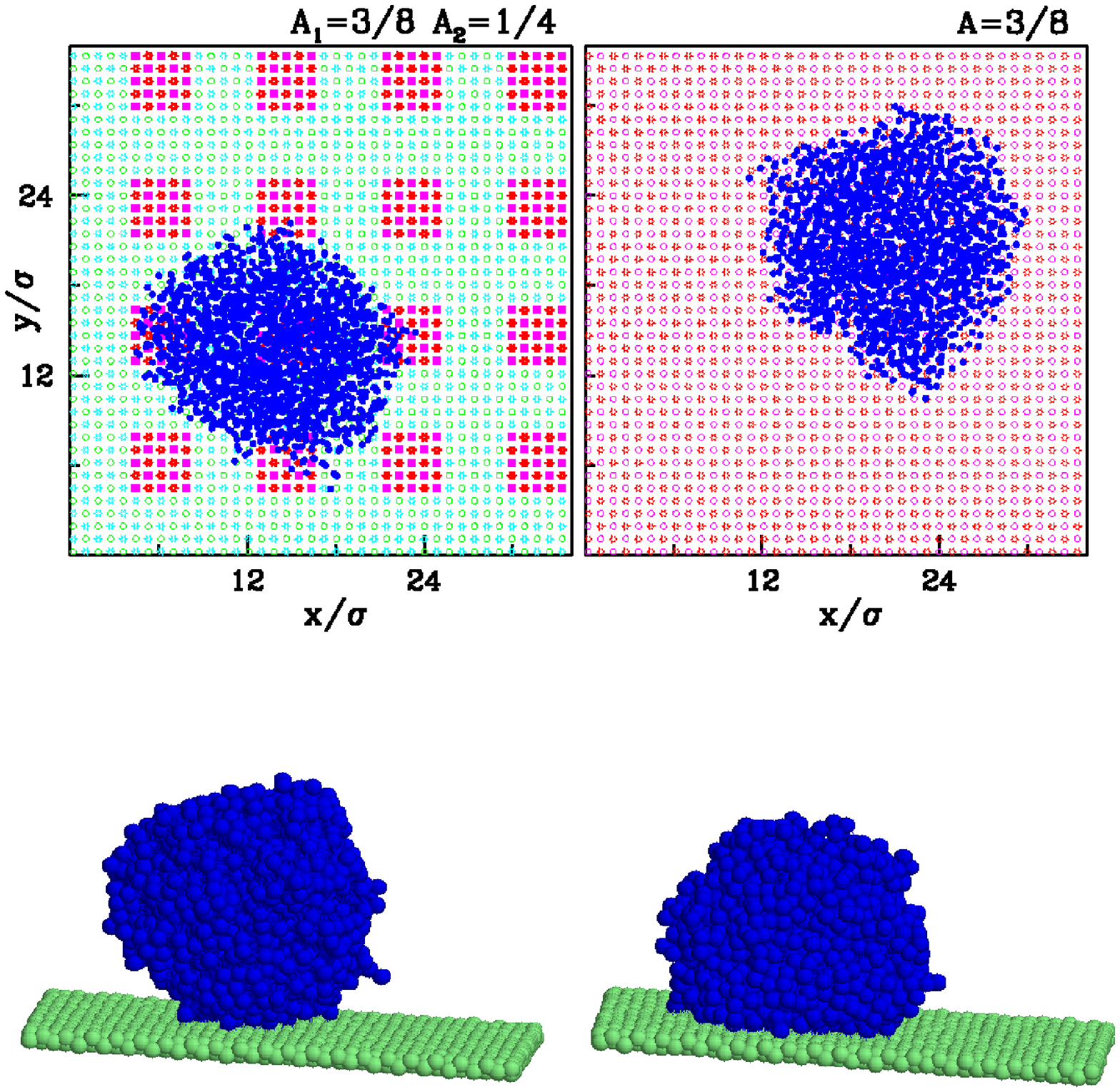}}
\vspace*{-2cm}
\caption{ }
\end{figure*}

\clearpage

\begin{figure*}
\epsfxsize=4.2in
\centerline{\epsffile{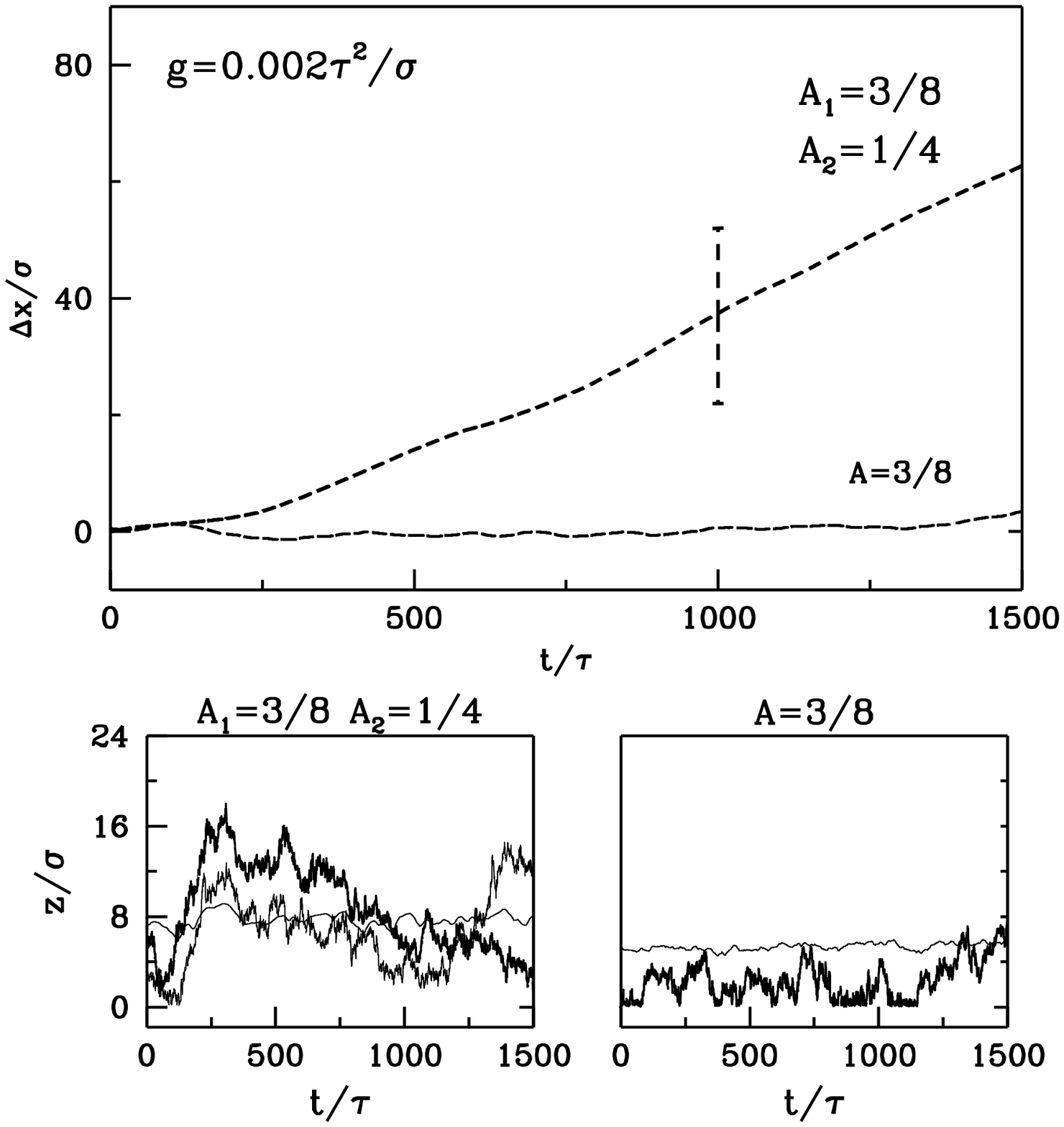}}
\caption{ }
\end{figure*}

\end{document}